# MammoGrid: A Service Oriented Architecture based Medical Grid Application

*Salvator Roberto Amendolia[2], Florida Estrella[1], Waseem Hassan[1], Tamas Hauer[1,2], David Manset[1,2], Richard McClatchey[1], Dmitry Rogulin[1,2] & Tony Solomonides[1]*

[1]*CCCS Research Centre, University of the West of England, Frenchay, Bristol BS16 1QY*
[2]*ETT Division, CERN, 1211 Geneva 23, Switzerland*

**Abstract**

The **MammoGrid** project has recently delivered its first proof-of-concept prototype using a Service-Oriented Architecture (SOA)-based Grid application to enable distributed computing spanning national borders. The underlying AliEn Grid infrastructure has been selected because of its practicality and because of its emergence as a potential open source standards-based solution for managing and coordinating distributed resources. The resultant prototype is expected to harness the use of huge amounts of medical image data to perform epidemiological studies, advanced image processing, radiographic education and ultimately, tele-diagnosis over communities of medical *'virtual organisations'*. The MammoGrid prototype comprises a high-quality clinician visualization workstation used for data acquisition and inspection, a DICOM-compliant interface to a set of medical services (annotation, security, image analysis, data storage and querying services) residing on a so-called 'Grid-box' and secure access to a network of other Grid-boxes connected through Grid middleware. This paper outlines the MammoGrid approach in managing a federation of Grid-connected mammography databases in the context of the recently delivered prototype and will also describe the next phase of prototyping.

## 1. Introduction

The aim of the MammoGrid project is to deliver a set of evolutionary prototypes to demonstrate that 'mammogram analysts', specialist radiologists working in breast cancer screening, can use a Grid information infrastructure to resolve common image analysis problems. The design philosophy adopted in the MammoGrid project concentrates on the delivery of a set of services that addresses user requirements for distributed and collaborative mammogram analysis. By moving towards a service-oriented architecture and supplying MammoGrid services interfaced to a Grid hosting environment through an Open Grid Services Architecture (OSGA) interface, the project aims to provide a path towards future Grid compliance. Inside the course of requirements analysis (see [1] and [2] for details) a hardware/software design study was also undertaken together with a rigorous study of Grid software available from other concurrent projects (now available in [3]). The MammoGrid project decided to adopt a lightweight Grid middleware solution (called AliEn (**Ali**ce **En**vironment) [4]) since the first OGSA-compliant Globus-based systems were yet to prove their applicability. The project delivers a set of medical MammoGrid services through a series of prototypes following a Service Oriented Architecture (SOA) philosophy and with an OGSA interface so that when suitable OGSA-based Grid solutions become available the MammoGrid services would be able to interface with those hosting environments.

AliEn [4] is a Grid framework developed to satisfy the needs of the ALICE high energy physics experiment at CERN for large scale distributed computing in the preparation phase of the experiment. It is built on top of the latest Internet standards for information exchange and authentication (SOAP, SASL, PKI) and common Open Source components (such as Globus/GSI, OpenSSL, OpenLDAP, SOAPLite, MySQL, CPAN). AliEn provides a virtual file catalogue that allows transparent access to distributed data-sets and provides top to bottom implementation of a lightweight Grid applicable to cases where handling of a very large number of files is needed. AliEn provides an insulation layer between different Grid implementations and a stable user and application interface to its community of AliEn users over extended timescales. As progress is being made in the definition of new Grid standards and interoperability, AliEn has been selected as a major component of the middleware for the upcoming EU funded EGEE [5] project, as discussed in a later section of this paper.

In the deployment phase, the CERN AliEn middleware has been installed and configured on a set of novel 'Gridboxes', secure hardware units which are meant to act as each hospital's single point of entry onto the MammoGrid. These units are configured and tested at CERN and Oxford, for later testing and integration with other Grid-boxes at hospitals in Udine (Italy) and Cambridge (UK). AliEn provides a software interface to a stack of Grid-compliant software layers, based on technologies emerging from the DataGrid project [6], among other Grid initiatives. As the MammoGrid project developed and new layers of Grid functionalities became available, AliEn facilitated the incorporation of new stable versions of Grid software in a manner that catered for controlled system evolution and provided a rapidly available lightweight but highly functional Grid architecture for MammoGrid. It is envisaged that eventually the AliEn code will evolve into EGEE and that the MammoGrid Grid application will experience a seamless transition between middleware platforms.

Other work in this area includes the NDMA [7] project in the US and the eDiamond [8] project in the UK. Our approach shares many similarities with these projects, but in the case of the NDMA project (one of whose principal aims is to encourage the adoption of digital mammography in the USA) its database is implemented in IBM's DB2 on a single server - that is, it avoids the technical issues of constructing a distributed database that exploits the emerging potential of the Grid. The MammoGrid project federates multiple (potentially heterogeneous) databases as its data store(s). MammoGrid is complementary to eDiamond and addresses different objectives: MammoGrid concentrates on the use of open source Grid solutions to perform epidemiological and CADe studies and incorporates pan-European data whereas eDiamond uses IBM supplied Grid solution to enable applications like 'find-one-like-it' for images and teaching studies on UK data samples. The current status of MammoGrid is that a single 'virtual organization' (VO) AliEn solution has been demonstrated using the so-called MammoGrid Information Infrastructure (MII) and images have been accessed and transferred between hospitals in the UK and Italy. The next stage is to provide rich metadata structures and a distributed database in a multi virtual organisation environment to enable epidemiological queries to be serviced and the implementation of a service-oriented (OGSA-compliant) architecture for the MII.

The MammoGrid project is funded by the EU $5^{th}$ Framework Programme, and is an example of an emerging specialised sub-category of e-Science projects—e-Health projects. One of the secondary aims of the MammoGrid project is to deliver generic solutions to the problems faced during development and deployment, both technical and social. This paper is presented in that spirit—we feel that the issues highlighted regarding VO Management for the MammoGrid project will be of interest not only to other mammography-related projects, but also to a wide variety of applications in which Grid middleware is utilized. The structure of this paper is as follows. In Section 2 we provide an overview of MammoGrid prototypes (P1 and P2). Next, in Section 3 we discuss the advantages of P2 over P1. In Section 4 we describe the P2 design in the context of EGEE project. Finally, in Section 5 we draw conclusions and indicate some potential areas for future work.

## 2. Description of MammoGrid Prototypes architecture

### 2.1 Prototype P1

The initial prototype deliverables included a mammogram image acquisition workstation and Standard Mammogram Format (SMF) algorithm [9], an installed AliEn-based Grid hosting environment [10], a configured 'Grid-Box', an optimised CALMA computer aided detection (CADe) algorithm for microcalcifications, plus user requirements and technical specification documents. An initial prototype P1 to be delivered as a consequence of all these activities aimed to demonstrate the MammoGrid workstation in a single 'Virtual Organisation' (VO) implementation of AliEn (i.e. centralised and simplified metadata and data file handling). This allowed testing of the MammoGrid front end with a *centralised AliEn architecture.*

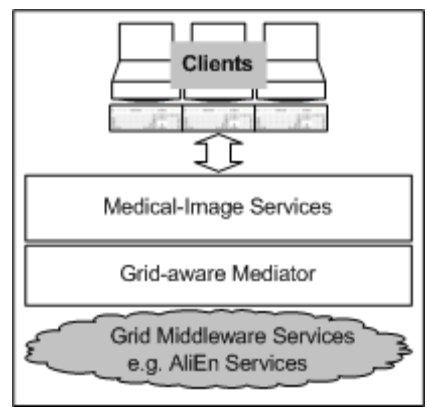

Figure 1 Vertical Stack of Grid Services

### 2.1.1 Prototype P1 Architecture

The MammoGrid prototype P1 enabled mammograms to be acquired into files that are distributed across Grid-boxes in the users' hospitals and for simple queries against those files to be executed. The mammogram images were transferred to the Grid-Boxes in DICOM [11] format where the AliEn services could be invoked to manage the file catalog and to deal with queries. Each Grid-Box is aware of the AliEn network of data files, managed by MySQL databases. In general, the medical image management on the Grid uses a vertical stack of Grid services (from here on referred to as 'services'). This is illustrated in Figure 1. The services are arranged in layers that build upon one another. The services are 'orchestrated' in terms of service interactions – how services are discovered, how services are invoked, what can be invoked, the sequence of service invocations, and who can execute.

The medical image (MI) services are directly invoked by authenticated MammoGrid clients. They provide a generic framework for managing image and patient data. The digitized images are imported and stored in DICOM [11] format. Relevant MI services include:

1. **add** for uploading DICOM files
2. **retrieve** for downloading DICOM files
3. **update** for updating contents of DICOM files
4. **query** for translating and executing a query on patient data
5. **addAlgorithm** for loading executable code
6. **executeAlgorithm** for executing an algorithm
7. **authenticate** for logging into the system.

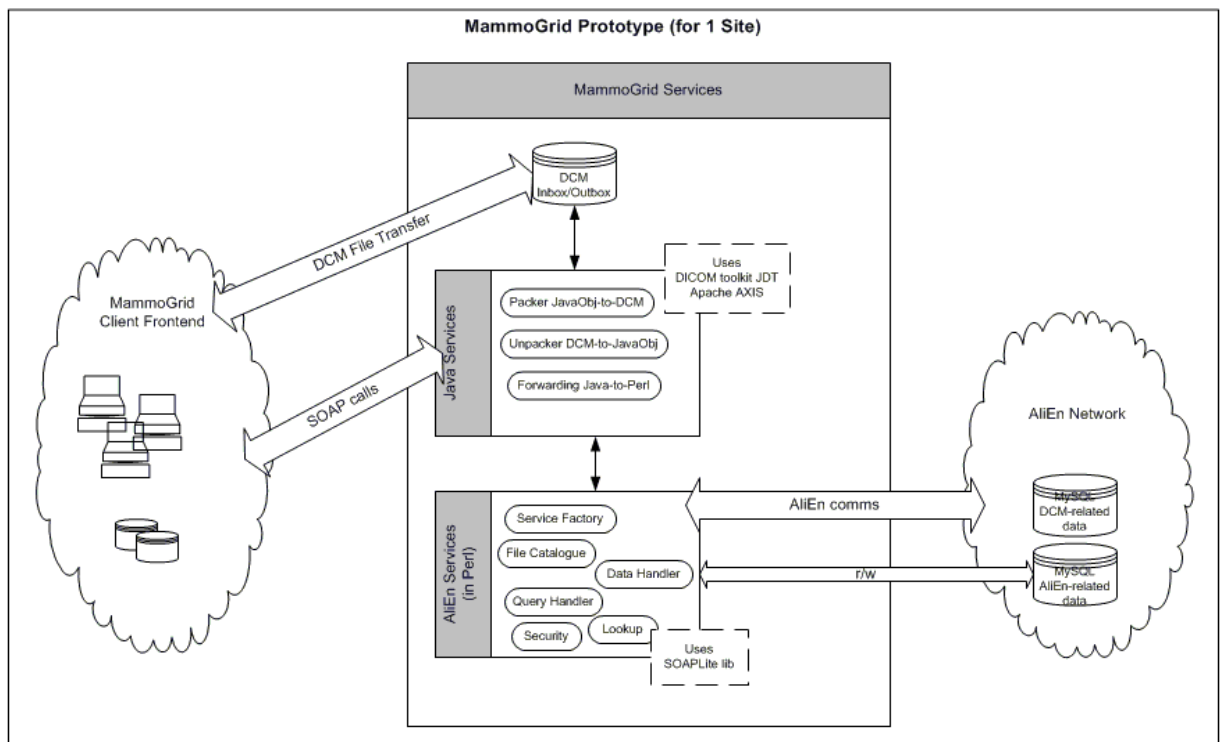

Figure 2: The MammoGrid prototype 1 architecture

Figure 2 shows the block diagram of the MammoGrid P1 architecture: the MammoGrid front end clinician workstation with a DICOM interface to the MammoGrid prototype Web Services and the Alien Perl-based middleware network. There are in fact two sets of Services; one, which is Java-based and comprises the business logic related to MammoGrid services and the other, Perl-based, which is the AliEn specific services for Grid middleware. As this architecture is Web Services-based, SOAP messages are exchanged between different layers. RPC calls are exchanged from Java specific services to Alien specific services. AliEn is also a Web Services-based solution being built on open source SOAP::Lite Perl module available from [12].

The following is the sequence (shown diagrammatically in Figure 3) in which messages are exchanged between the different MammoGrid components depicted in Figure 2. A client from the MammoGrid client frontend connects to the Java Portal through an Authentication module (see 1 in the Figure 3) with the message transfer being through SOAP [doc/literal]. The Authentication module in turn sends the call to a Secure Factory (see 2) for client authentication. After successful authentication, an Interface instance is created for this particular client (see 3). The URL of the Interface is sent back to the Authentication module (4), which in turn creates a Portal instance for the client (see 5). The URL of the portal is sent back to the MammoGrid client (see 6) to enable message exchange between the client and the Portal (e.g. 7). Messages are passed from the Portal to the Perl Interface (see 8), which in turn connects to the Grid network. From the Grid Network a response arrives which depends on the type of call from the client (see 9) and this response is finally sent back to the MammoGrid client (see 10).

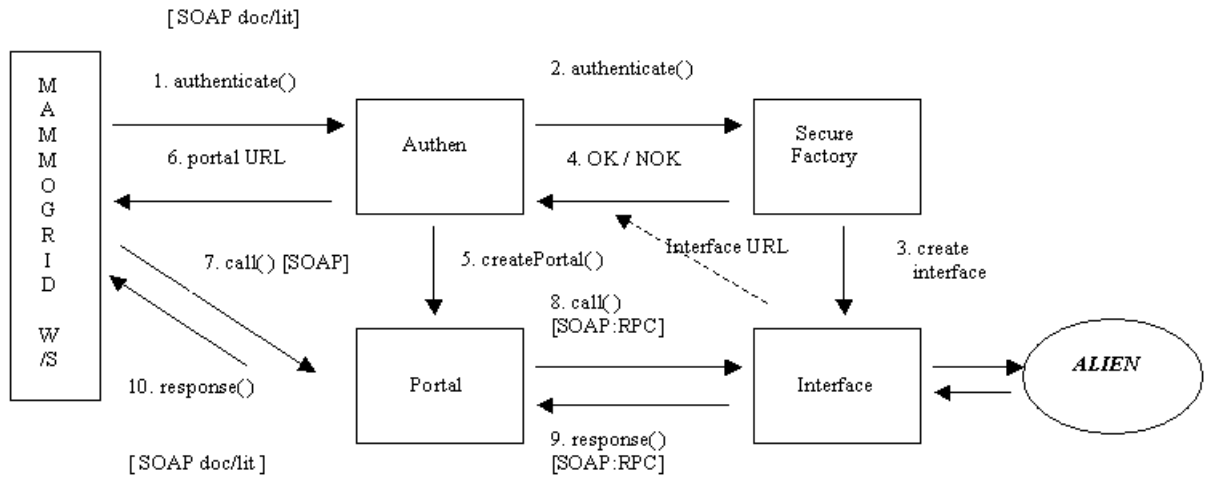

Figure 3: Message exchange between various components

### 2.1.2 The MammoGrid P1 Services

The following is a brief description of the MammoGrid P1 services, which are invoked in the middleware:

1. **Database Proxy:** To avoid linking with any specific database driver on the client side, all connections to the databases are channelled via a Proxy Service. The application connects to the Proxy Service by means of a special AliEnProxy driver so that the real database driver and libraries need to be installed only in the place where an instance of the database Proxy Service is running. The Proxy Service also acts as an Authentication Service for a certificate-based authentication.
2. **Configuration Manager:** The Configuration Manager is responsible for discovery and read-only interactions with the LDAP server. It extracts all relevant configuration parameters that apply to a particular VO in the context of a site and a specific host. The information is kept in a cache to avoid frequent LDAP lookups.
3. **Authentication**: The Authentication Service is responsible for checking users' credentials. AliEn uses the SASL protocol for authentication and implements several SASL mechanisms (GSSAPI using Globus/GSI, AFS password, SSH key, X509 certificates and AliEn tokens).
4. **Brokers:** AliEn jobs use the Condor ClassAds as a Job Description Language (JDL). When users submit a job to the system it enters into the task queue and the Resource Broker becomes responsible for it. The Broker analyses job requirements in terms of requested input files, requirements on execution nodes, and job output.
5. **File Transfer:** This service typically runs on the same host as the Storage Element and provides the scheduled file transfer functionality. The File Transfer Daemons (FTD) are mutually authenticated using certificates and will perform file transfer on users' behalf.
6. **Optimizers:** While the jobs or file transfer requests are waiting in the task queue, the Job and Transfer Optimizers inspect JDL scripts and try to fulfil requests and resolve conflicts.
7. **Storage Elements (SE):** The Storage Element (SE) is responsible for saving and retrieving the files to and from the local storage. It manages disk space for files and maintains the cache for temporary files.
8. **Computing Elements (CE):** The Computing Element (CE) is an interface to the local batch system. At present, AliEn provides interfaces to LSF, PBS, BQS, DQS, Globus and Condor. The task of a CE is to get hold of jobs JDLs from the CPU Server, translate them to the syntax appropriate for the local batch system syntax and execute them.
9. **The Managers (Transfer and Job Manager):** take care of the requests concerning tasks. They insert the tasks to be done in a queue, and report the status of the tasks. They do not assign the tasks to the resources as this is done by the brokers.
10. **The Factory:** is a service that can start several instances of the Portal Service for a given user.
11. **MammoGrid-specific high-level services:** are the ones, which are interfacing with the MammoGrid client frontend and the AliEn middleware. The MammoGrid services are written in Java and use DICOM toolkits, among others, to handle data coming from the MammoGrid client.

The Grid middleware services address issues related to resource management, information discovery and security infrastructure. These services are expected to be OGSA-compliant to ensure interoperability between the MammoGrid and other Grid implementations and a migration path to fully tested and robust Grid as these standards become available. In P1 architecture all the above service invocation are done in the context of a single VO set up. This means there is one VO called "MammoGrid" and each hospital represents a single site inside the VO. This will be explained in a greater detail in the next section.

### 2.1.3 The Prototype P1 deployment

The deployment of P1 is based on a Central Node (CN at CERN) which is used to hold all AliEn metadata and its file catalog and a set of Local Nodes (LNs at Oxford, Cambridge and Udine) on which clinicians capture and examine mammograms using the clinician workstation interface. The list of services which should be running on the CN and on the LNs is shown in Figure 4. These services include both the AliEn services (implemented in Perl) and the MammoGrid services (implemented in Java). The CN and LNs are connected through a VPN router.

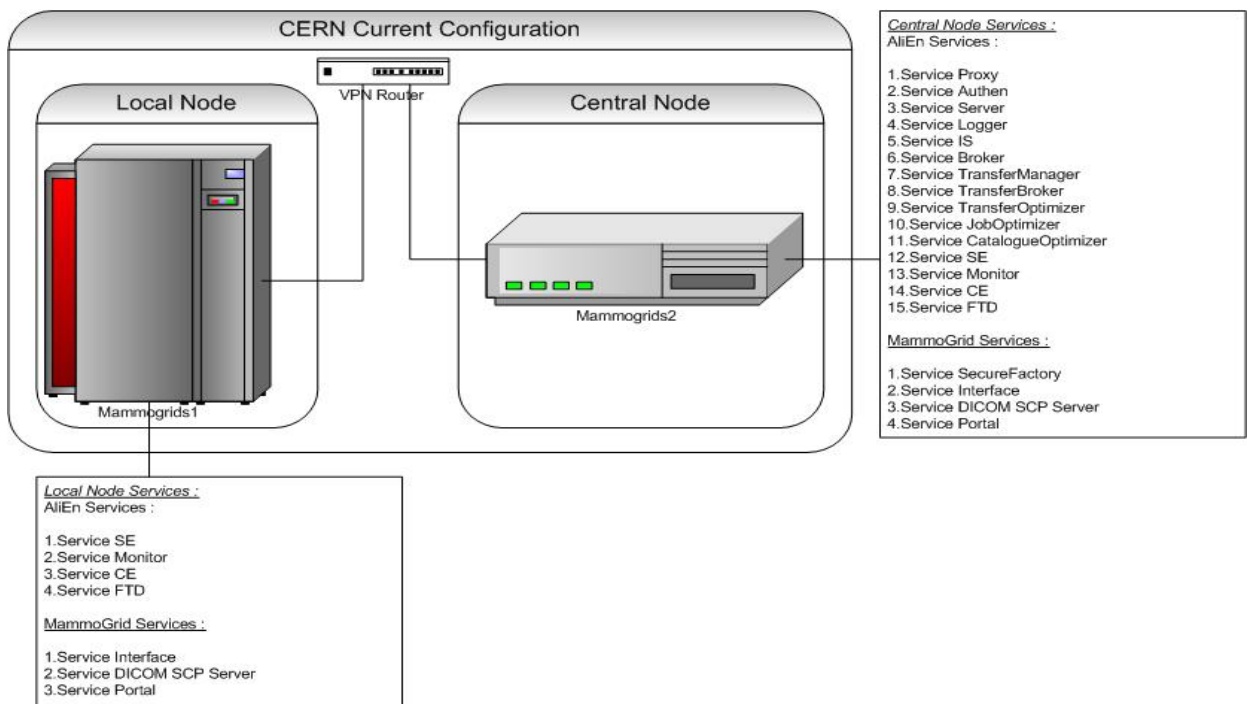

Figure 4: The P1 physical configuration

In essence, the MammoGrid prototype P1 has the following characteristics:

1. The prototype is based on SOA, i.e. a collection of coordinated services. There are two types of services – application services and middleware services (see Figure 4 in which MammoGrid Services are the application services and AliEn services are the middleware services).
2. This prototype architecture is for a single VO, which encompasses the Cambridge, Udine, and Oxford sites and a central server site.
3. Each local site has a MammoGrid grid box. The local grid box contains the MammoGrid high level services and local AliEn services. The central grid box contains the AliEn central services.
4. The Grid hosting environment (GHE) is AliEn. AliEn provides local middleware services and central middleware services. AliEn manages the Grid infrastructure and AliEn services are written in Perl.

### 2.1.4 Analysis and Discussion of the P1 architecture

#### 2.1.4a Federation in a Single Virtual Organisation (Centralized Architecture):

The resources in the federation, e.g. hospitals, research institutes, universities, are governed by the same sharing rules with respect to authentication, authorization, and resource and data access. These rules create a highly controlled environment, which dictates what is shared, who is allowed to share, and the conditions under which sharing occurs among medical sites. Federation in this application implies cooperation of independent medical sites. Individually, these sites are autonomous in that they have separate and independent control of their local data. Collectively, they participate in a federation, and the federation is governed by the organization.

Figure 5 illustrates an AliEn-connected single VO configuration. The AliEn middleware provides services (e.g. authentication, data access, resource broker, file transfer) that facilitate the management of resources in the VO. In essence, the medical community dictates the interaction protocol and AliEn implements and enforces these rules on the participating entities of the organization through services. For detailed discussion, see the MammoGrid Technical Specification in [13]. The medical sites in a single VO operate within the rules specified by a governing organization. In reality, there are many (co-)existing organizations, with different rules and protocols. Typically, hospitals have different regulations and governments have different legislations. A federation of multiple VOs extends the single VO set up by inter-connecting potentially disparate VOs.

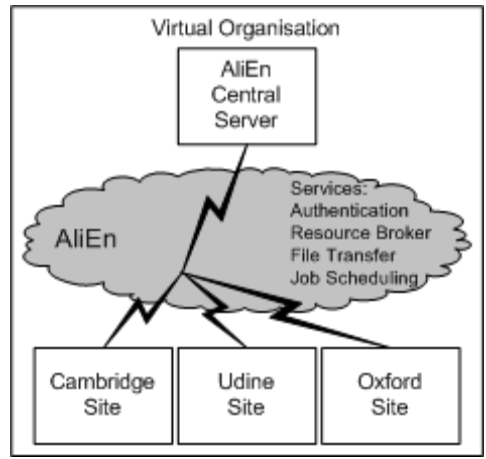

Figure 5: Federation in an AliEn-connected VO

**2.1.4b Web Services based architecture:**

The overall architecture in MammoGrid P1 is Web Services (WS) based. WSs are stateless and persistent, where data is not retained among invocations and services outlive their clients. OGSA defines Grid services as extending Web services and are stateful and time-limited instances with delegated authentication credentials. Thus a Grid-based SOA is a coordinated diverse set of Grid services in a virtual organisation. OGSA draws on the same infrastructure as in WSs – XML, SOAP, and WSDL. In other words OGSA enhances Web Services to accommodate requirements of the Grid. It is reasonable to expect that in the future Grid applications will be required to be OGSA-compliant [14]. The fundamental concept behind OGSA is that it is a service-oriented Grid architecture powered by Grid services [15]. A Grid service is a special Web service [16] that provides a set of well-defined interfaces that follow specific conventions [17].

Despite the fact that OGSA represents a long-overdue effort to define a Grid architecture, it is a relatively new standard [15]. The Open Grid Service Infrastructure (OGSI) was the first set of formal and technical specifications of the concepts described in OGSA, but many problems were reported regarding these. In order to circumvent the discrepancies in the OGSI specifications a new standard is emerging, which is called Web Services Resource Framework (WSRF) [18]. WSRF represents a refactoring and evolution of OGSI that delivers essentially the same capabilities in a manner that is more in alignment with the Web Services [16] community. The most valuable aspect of WSRF is that it effectively completes the convergence of the Web services and Grid computing communities.

WSRF specifications build directly on core Web services standards, in particular WSDL, SOAP and XML, and exploit capabilities provided by WS-Addressing [19]. WSRF represents a refactoring and evolution of OGSI that delivers essentially the same capabilities in a manner that is more in alignment with the Web Services community. As such, it represents an important next step towards the larger goal of a comprehensive Open Grid Services Architecture that supports on-demand, utility computing, collaborative and other Grid scenarios within a Web services setting. Since the MammoGrid current architecture is Web Services based it can be integrated with anything based on WSRF standard.

**2.1.4c Compatibility with other middleware:**

The current prototype is based on the AliEn middleware, which is a Web Services based architecture. At the moment there are several different Grid implementations available with different hosting environments. These heterogeneous hosting environments are not interoperable with one another. Consequently there is a measure of incompatibility between different middleware. In this regard an effort has been made very recently to interoper-ate with EDG middleware by the AliEn team [20]. In future there will be a need to bring OGSA/WSRF-compliance into AliEn to ensure future Grid compatibility.

## 2.2 Prototype P2

As timely delivery of a Grid-compliant architecture for MammoGrid was crucial to the project, the implementation of the prototypes were planned to be delivered in phases. The purpose of P1 was mainly to exploit the existing capabilities of the technologies involved. Prototype P2 aims to have distributed metadata on each deployed Grid-Box and *multiple VOs*, one at each of CERN, Oxford, Udine and Cambridge. In general each Grid-Box will have three major components:

1. MammoGrid services including a DICOM front-end which interfaces with the MammoGrid workstation to save and retrieve local mammograms according to the clinicians' queries.
2. A set of Services including metadata/query handler which interacts with the local metadata to resolve the queries [21] and
3. OGSA-compliant Grid services [22] which allow the VO to interact and exchange data with other VOs (e.g. the Udine VO sharing access to the Cambridge VO mammogram data) i.e. a multi-VO infrastructure.

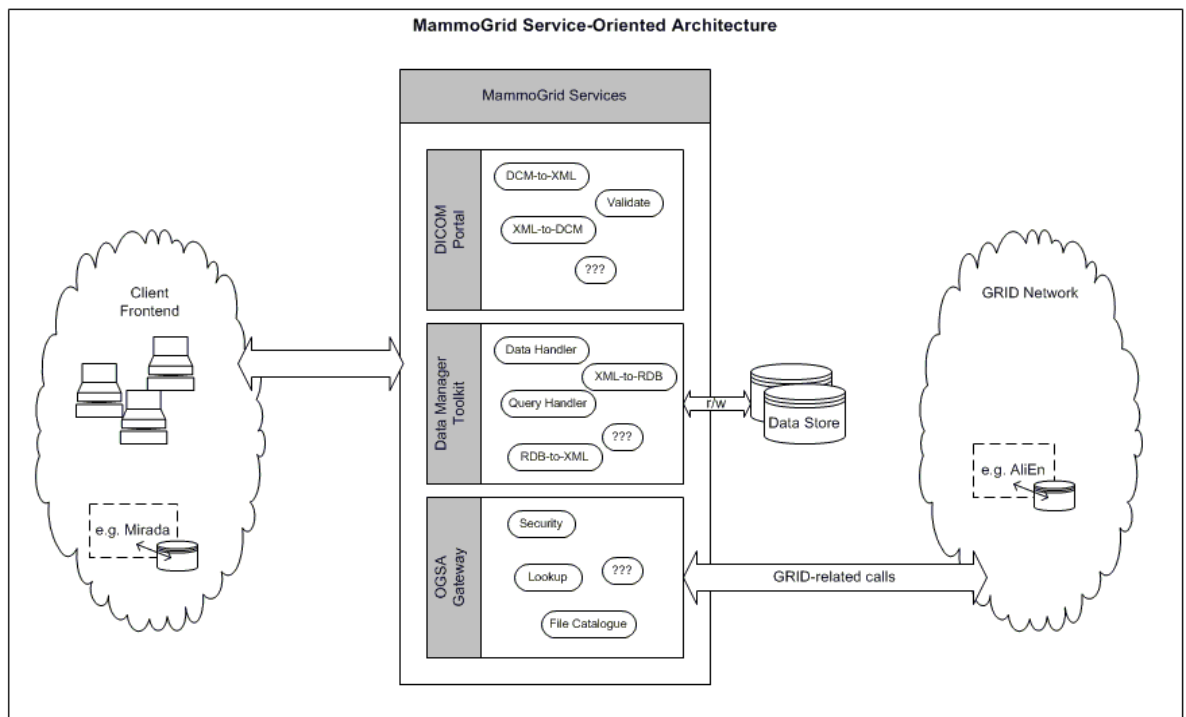

Figure 6: The MammoGrid prototype 2 architecture.

### 2.2.1 Prototype 2 Architecture

P2, which will be delivered in the final months of the project, will be based on a Service-Oriented Architecture with clear portals/gateway/interfaces both to external systems such as the CADe system, the MammoGrid acquisition hardware and to the Grid network. It is diagrammatically shown in Figure 6 and comprises a set of MammoGrid Services including a DICOM portal facilitating the exchange of information between the MammoGrid workstation and the Grid-Box. This includes validation services and DICOM-to-XML and XML-to-DICOM translators, as well as an OGSA Gateway consisting of Security, Look-up and File Catalogue services. Hence the DICOM client front-end will enable any DICOM-compliant external device to exchange information via a set of MammoGrid Services with any OGSA back-end Grid network.

In view of the upcoming EGEE middleware, it is expected that the next prototype of MammoGrid Project will be based on the new middleware. For further details see section 5. Note that the P1 architecture had a tight coupling between different layers of services but P2 architecture will provide loose coupling between different vertical layers of services. P2 architecture will be multi-VO based; its details are given in sub-section 2.2.3.

### 2.2.2 The P2 Architecture deployment

The prototype P2 deployment is shown in Figure 7; P2 activities revolve around creating a multi-VO architecture. Figure 7 illustrates this configuration. In essence, the P2 architecture aims to exhibit the following:

1. All services – MammoGrid high level services and middleware services – are OGSA-compliant services. The MammoGrid services will be re-designed, re-developed and re-deployed as OGSA-compliant Grid services. It is anticipated that AliEn (and now the EGEE middleware) and other Grid middleware implementations will move towards compliance in the near future.

2. A set of OGSA-compliant abstract level services will be added to serve as mediator between the MammoGrid services and the middleware services. With middleware implementations still in their infancy, these abstract services will (1) remove the dependency on Grid-related activities (e.g. new technologies, new implementations, additional OGSA protocols), (2) allow the use of different Grid implementations as and when they become available and usable, and (3) create a modular approach in the design and development of the MammoGrid application which can sit on top of any OGSA-compliant Grid.

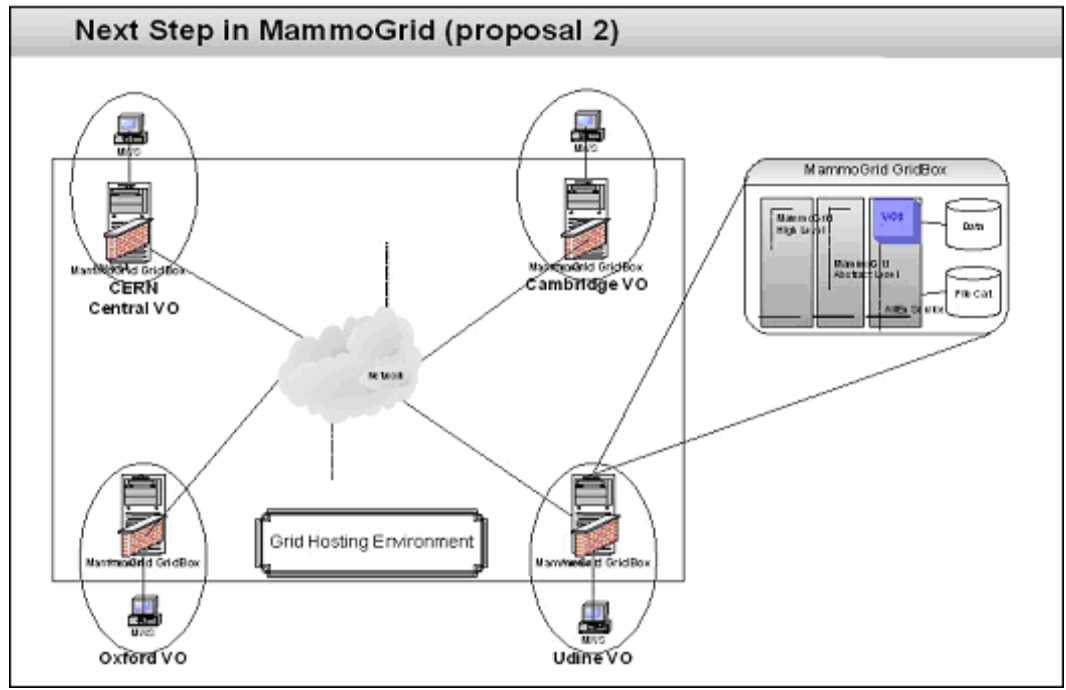

Figure 7: The MammoGrid P2 prototype

3. Each hospital will belong to a virtual organisation. This implies three VOs for the three hospitals. Each VO will maintain its own access and authentication rules. Each VO will maintain its own data. A fourth VO, the central VO, will be responsible for inter-VO communication. A set of inter-VO-related services (currently referred to as VOMS – Virtual Organisation Management Services) is anticipated in future enhancements of Grid implementations.

4. AliEn middleware services will include Virtual Organisation Services (VOS). These services will be in charge of authentication and authorization between Virtual Organizations. In case a user from one VO has to access resources of another allowed VO, the VOS of both VOs will check authorization and create the appropriate credentials.

### 2.2.3 Federation in a Multiple Virtual Organisation

The medical sites in a single VO operate within the rules specified by a governing organization. In reality, there are many (co-)existing organizations, with different rules and protocols. Typically, hospitals have different regulations and governments have different legislations. A federation of multiple VOs extends the single VO set up by inter-connecting potentially disparate VOs (see Figure 8). However, this VO mechanism is not sufficient to guarantee security and privacy. Some technical additions are needed to meet security requirements. Thus, two security aspects must be considered, first data encryption across Grid communications and second user authentication and authorization across VOs.

In order to preserve privacy, patient personal data is – as a first step – partially encrypted in MammoGrid to facilitate anonymization. As a second step, when data is transferred from one service to another through the network, communications are established through a secure protocol HTTPS with encryption at a lower level. Each Grid-box is in fact made part of a VPN (Virtual Private Network) in which allowed participants are identified by a unique host certificate.

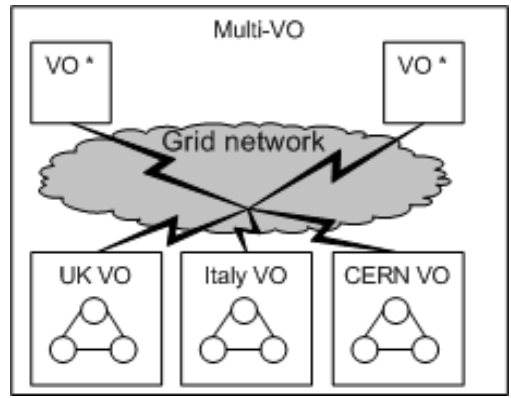

Figure 8: Multi-VO Configuration

The general requirements for multi-VO (security) architecture are listed below (adopted from [23]):

1. **Scalability:** Ability to manage a steady increase in users and resources as collaborations between other organizations (hospitals) grow.
2. **Manageable and maintainable:** Adding, removing and modifying user privileges need to be kept easy and intuitive.
3. **Under the control of the resource end:** Organizations (hospitals) are required to have complete access control over their data.
4. **Minimum intervention at the Data Layer:** To keep security issues to a minimum.
5. **Ability to utilize existing Access Control Models:** Resource providers have existing access control mechanisms that are reliable and proven.
6. **Future integration capabilities** with other Grid related applications.

The current AliEn design uses a hierarchical LDAP database to describe the static configuration for each VO. This includes People, Roles, Packages, Sites and Grid Partitions as well as the description and configuration of all services on remote sites. The code that is deployed on remote sites or user workstations does not require any specific VO configuration files, everything is retrieved from the LDAP configuration server at run time thus allowing user to select VO dynamically.

Due to weak coupling between the resources and the Resource Brokers in the AliEn Grid model it is possible to imagine a hierarchical Grid structure that spans multiple AliEn and "foreign" Grid but also includes all resources under the direct control of one top level VO called the "Super VO". The connectivity lines in Figure 9 above represent the collaboration and trust relationships. In this picture the entire foreign Grid can be represented as a single Computing and Storage Element (albeit a potentially powerful one). In this manner an AliEn-based VO can interoperate with other VOs (such as the European Data Grid EDG [24], or other GT3-based VOs). Along the same lines, an AliEn-AliEn interface allows the creation of a federation of collaborating Grids. The resources in this picture can be still shared between various top level VOs according to the local site policy so that Grid federations can overlap at resource level. However, when it comes to implementing a data Grid suitable for medical applications, this kind of flexibility comes at a price – each VO must maintain its own File and Metadata Catalogue. Figure 9 shows an AliEn multi-VO architecture.

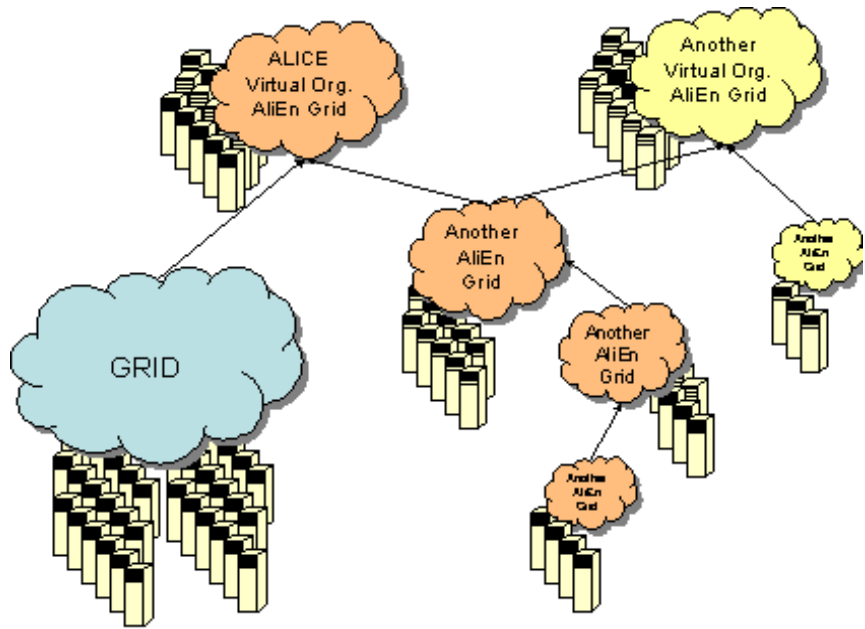

Figure 9: AliEn in a multiple Virtual Organisation environment.

## 3. Comparison of the P1 and P2 architectures

The MammoGrid requirements analysis process revealed a need for hospitals to be autonomous and to have 'ownership' of all local medical data. As outlined above, the P1 design is based on a single-VO architecture, implying a partial compromise over the confidentiality of patients' data. In this architecture all metadata is centralized on one site and this requires the copying of some summary data from the local hospital to that site in order to satisfy the predicates required for query resolution. Practically this is not feasible other than in specialist research studies where appropriate authority for limited replication of medical data between hospitals can be sought and granted. Since the P2 design will be based on a multi-VO architecture where there will be no centralization of data, all metadata will be managed alongside the local medical data records and queries will be resolved at the hospitals where the local data are under governance. This will provide a more realistic clinical solution, which matches more closely the legal constraints imposed by governance of patient data. With the incorporation of a multi-VO set up the overall metadata will become truly federated and will provide a more secure solution for the medical community.

Furthermore, full adaptation of the Grid philosophy suggests that the federation, (normally regarded as just independence and autonomy) should not be confined to the database level but should be realized in all aspects of Grid services (including areas such as authentication, authorization, file replication, etc.). The solution, which will be adopted in P2 design, is a federation of VOs (possibly hierarchical in structure), where the boundaries of the organizations naturally define the access rights and protocols for data-flow and service access between 'islands' that provide and consume medical information.

As the P1 design was based on simple Web Services, interoperability with other middleware was not practically possible. This is mainly due to the fact that different Grid hosting environments require different underpinning technologies; the lack of common communication protocols results in incompatibilities. While OGSA has adopted a service-oriented approach to defining the Grid architecture, it says nothing about the technologies used to implement the required services and their specific characteristics. That is the task of OGSI/WSRF. The WSRF working group opted to build the Grid infrastructure on top of Web services standards, hence leveraging the substantial effort, in terms of tools and support that industry has been putting into the field. Since the prototype P1 was designed to demonstrate and benchmark client-related and middleware-related functionalities, the service access design (API) was kept as simple as possible using a handful of web-service definitions. The focus of the next MammoGrid milestone will be the demonstration of Grid services functionality and to that end it is planned to implement an (OGSA-compliant) Grid-services based infrastructure. As a result the focus will be on interoperability with other OGSA-compliant Grid services. It is observed that the major trend in Grid computing today is moving closer to true web-services (from OGSI) and we expect a convergence between different Grid protocols in time for the completion of our existing web-services design.

Finally, the P1 architecture is rather tightly coupled and it is intended that the next architecture will be loosely coupled. Tight coupling is an undesirable feature for Grid-enabled software, which should be flexible and inter-operable with other Grid service providers. Thus a key requirement of the second prototype is to conform fully to a loosely coupled paradigm. In this spirit it is also necessary to clearly separate the data layer from high-level functionality so that the security requirements can be defined transparently. In other words it is necessary to have minimum intervention at the data layer to keep the points of security consideration as low as possible.

## 4. Future Work – a P2 Design that is EGEE based

At the outset of the MammoGrid project there were no demonstrable working alternatives to the AliEn middleware and it was selected to be the basis of a rapidly produced MammoGrid prototype. Since then AliEn has steadily evolved (as have other existing middleware) and the MammoGrid project had to cope with these rapidly evolving implementations. Currently AliEn is being considered as the basis of middleware for numerous applications. For example during the early analysis of ARDA (A Realisation of Distributed Analysis) [25] initiative at CERN, AliEn was found to be the most stable and reliable middleware during the production phases of ALICE experiment.

At the same time, work has been started in a new EU-funded project called EGEE (Enabling Grid Environment for E-science). EGEE is a follow-up project to the recently concluded EU-funded EDG (European Data Grid) project. The aim of EGEE is to build on the recent advances in Grid technology and develop a Grid infrastructure across Europe that will be constantly available on a round-the-clock basis. The project will primarily concentrate on three core areas. The first area is to build a consistent, robust and secure Grid network. The second area is to continuously improve and maintain the middleware in order to deliver a reliable service to users. The third area is to attract new users from industry as well as science and ensure they receive the high standard of training and support they need. The Grid will be built on the EU Research Network GEANT [26] and exploit Grid expertise generated by many EU, national and international Grid projects to date.

The EGEE project community has been divided into 12 partner "federations", consisting of 70 partner institutions and covering a wide-range of both scientific and industrial applications. Two pilot areas have been selected to guide the implementation and certify the performance and functionality of the evolving infrastructure. One is the Large Hadron Collider Computing Grid (LCG), another is Biomedical Grid, where several communities are facing equally daunting challenges to cope with the flood of bioinformatics and healthcare data. EGEE based middleware is currently in development phase. The design of this new middleware revolves around the ideas of AliEn. According to some discussion with the middleware experts of EGEE middleware, the interface of the new middleware will be same as that of its predecessor i.e. AliEn. Although there will be several improvements in the underlying architecture the interface will essentially remain the same and it will be more or less seamless for projects like MammoGrid to become early adopters of the new EGEE middleware, as is planned for the P2 prototype.

## 5. Conclusion

The proliferation of information technology in medical sciences will undoubtedly continue, addressing clinical demands and providing increasing functionality. The MammoGrid project aims to advance deep inside this territory and explore the requirements of evidence-based, computation-aided radiology, as specified by medical scientists and practicing clinicians. Currently the MammoGrid project is undertaking the deployment and on-site testing of a first prototype in which a reduced set of mammograms are being tested between sites in the UK, Switzerland and Italy. Clinicians are being closely involved with these tests and it is intended that a subset of the clinician queries will be executed to solicit user feedback. Within the next few months a rigorous evaluation of the prototype will then indicate the usefulness of the Grid as a platform for distributed mammogram analysis and in particular for resolving clinicians' queries. In its first year, the MammoGrid project has faced interesting challenges originating from the interplay between medical and computer sciences and has witnessed the excitement of the user community whose expectations from the a new paradigm are understandably high. As the MammoGrid project moves into its final implementation and testing phase, further challenges are anticipated; this will test these ideas to the full.

In conclusion, this paper outlines the MammoGrid services-based approach in managing a federation of Grid-connected mammography databases in the context of the recently delivered phase 1 prototype and outlines the strategy being adopted for its deployment. The current status of MammoGrid is that a single 'virtual organisation' AliEn solution has been demonstrated using the MII and images have been accessed and transferred between hospitals in the UK and Italy. The next stage is to provide rich metadata structures and a distributed database in a multi virtual organisation environment to enable epidemiological queries to be serviced and the implementation of a service-oriented (OGSA-compliant) architecture for the MII. In doing so, the MammoGrid project is envisioning to become an early adopter of the new EGEE middleware, which will eventually be a seamless transition from the current Grid Middleware to the new EGEE middleware.

## Acknowledgements

The authors take this opportunity to acknowledge the support of their institutes and numerous colleagues for their contribution in different aspects of this research project.